\documentclass[aps,showpacs,twocolumn,amsmath,amsfonts,amssymb]{revtex4-1}
\usepackage{graphicx}
\usepackage[tight]{subfigure}
\usepackage{times}
\usepackage{color}
\usepackage{bm}
\usepackage{array}
\usepackage[colorlinks=true,bookmarks=false,citecolor=blue,linkcolor=red,urlcolor=blue]{hyperref}
\usepackage{appendix}

\begin{document}

\title{Thermoelectric response and entropy of fractional quantum Hall systems}
\author{D. N. Sheng$^1$ and Liang Fu$^2$}
\affiliation{
$^1$Department of Physics and Astronomy, California State University, Northridge, California 91330, USA\\
$^2$Department of Physics, Massachusetts Institute of Technology, Cambridge, Massachusetts 02139, USA
}

\begin{abstract}
We  study thermoelectric transport properties  of
fractional quantum Hall systems based on  exact diagonalization calculation.
Based on the relation between thermoelectric response and thermal entropy, we demonstrate that thermoelectric Hall conductivity $\alpha_{xy}$ has powerlaw scaling  $\alpha_{xy} \propto T^{\eta}$ for
gapless  composite Fermi-liquid states at filling number $\nu=1/2$ and $1/4$ at low temperature
($T$),  with exponent $\eta \sim 0.5$ distinctly different from Fermi liquids.
The powerlaw scaling remains unchanged for different forms of interaction including Coulomb and short-range ones,
demonstrating the robustness of  non-Fermi-liquid behavior at low $T$.
In contrast, for $1/3$ fractional quantum Hall  state,  $\alpha_{xy}$ vanishes at low $T$ with an activation gap associated with neutral collective modes rather than charged quasiparticles. Our results establish a new manifestation of the non-Fermi-liquid nature of quantum Hall fluids at finite temperature.
\end{abstract}

\date{\today}
\maketitle

{\it Introduction.---}
Thermoelectric phenomena that provide the direct conversion between heat and electricity  are interesting and useful. Decades of research has been devoted to finding materials and methods to increase thermoelectric energy conversion efficiency \cite{Goldsmid, Benenti2017}.
Recent theoretical works suggested the possibility of record-high thermoelectric conversion efficiency in semiconductors and semimetals under a quantizing magnetic field \cite{Skinner, Fu}, where thermoelectric response is directly related to entropy\cite{Obraztsov, Girvin1982, Bergman2009, Kozii}.
Based on this relation, it is found that thermopower of 3D Dirac/Weyl materials in the quantum limit increases unboundedly with the magnetic field  \cite{Skinner, Ong2017, ZhangExpt, Han2019, Viktor2019}. Very recently, it is shown that  two-dimensional  quantum Hall systems  can reach thermoelectric figure of merit on the order of unity down to low temperature ($T$), as a consequence of the thermal entropy from the massive Landau level (LL) degeneracy\cite{Fu}.

The degeneracy of a partially filled Landau level is lifted by disorder and electron-electron interaction. Therefore, thermoelectric response of quantum Hall systems is expected to depart from the noninteracting and clean limit when thermal energy $k_B T$ is smaller than disorder-induced Landau level broadening $\Gamma$ or a characteristic  energy scale proportional to electron interaction strength. Previous works\cite{Girvin1982, Zhu2010}  have shown that disorder leads to a $T$-linear thermoelectric Hall conductivity $\alpha_{xy} \propto T$ for $k_B T \ll \Gamma$, in accordance with thermal entropy of disorder broadened Landau levels.

On the other hand, in clean systems electron-electron interaction  lifts the massive Landau level degeneracy and forms the many-body ground state at fractional filling. These include gapped fractional quantum Hall (FQH) states\cite{Tsui1982} and gapless composite Fermi liquids\cite{Jain1989, Halperin1993}, which provide a fertile ground for exotic quantum states of matter. After nearly four decades of theoretical and experimental studies,  ground state properties and low-energy excitations at various fractional fillings are
largely understood. In contrast, much less is known about FQH  liquids at finite temperature. Based on the relation between the entropy and the thermoelectric transport coefficient,
a  linear $T$ scaling behavior for thermopower $S_{xx}$   has been conjectured \cite{Cooper1997}  for  composite Fermi liquid states.
While there are a few measurements\cite{Ying1994,Bayot1995,Chickering2013, Behnia2016, Checkelsky2009, Wei2009, Wang2010}, to our knowledge no  theoretical or numerical calculations on finite $T$ thermoelectric transport coefficients
for fractional quantum Hall states exists so far.

In this work, we investigate   thermoelectric Hall response for interacting quantum Hall systems
 through exact diagonalization calculations of  entropy of finite size systems at finite temperature.  For even denominator filling numbers $\nu=1/2$ and $1/4$, we identify robust power law scaling behavior
of the thermoelectric Hall conductivity $\alpha_{xy} \propto T^{\eta}$ in a wide temperature range, with the exponent $\eta \sim 0.5$ distinctly different
from the linear $T$ behavior of Fermi liquids.
We further show that the scaling behavior is  robust against weak disorder, and the exponent $\eta$ gradually increases with disorder strength. In contrast,  for 1/3 FQH system, we observe  a vanishing $\alpha_{xy}$  at low $T$  below the excitation  gap.
Our prediction of
anomalous power-law temperature dependence of thermoelectric response establishes a new fundamental property of $\nu=1/2$ and $\nu=1/4$ quantum Hall fluids, which can be measured in future experiments.

\textit{Model and Method.---}
We consider a  two dimensional (2D) electron system subject to a perpendicular strong magnetic field, whose energy spectrum is composed of discrete LLs. Throughout this work we assume the cyclotron energy is much larger than other energy scales set by interaction, disorder scattering or temperature, so that it suffices to work with the restricted Hilbert space of the partially
 filled  LL.

\begin{figure}[t]
\includegraphics[width=1.7\linewidth]{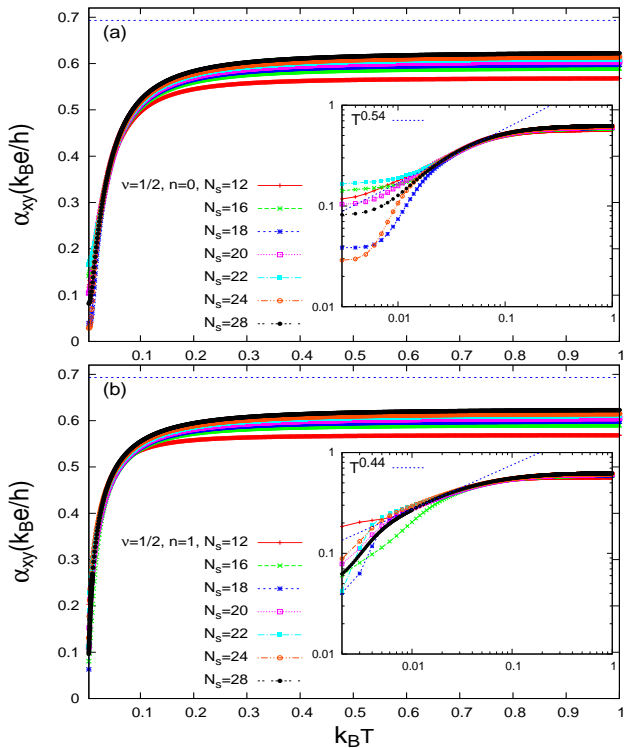}\\
\caption{Thermoelectric Hall coefficient  $\alpha_{xy}$ (in units of $k_Be/h$) for electrons
 at filling number $\nu=1/2$ of the n-th LL with  the number of flux quanta varying
between $N_s=12-28$. The thermal energy $k_BT$ is in units of energy $e^2/\epsilon\ell =1$ in all figures.
 $\alpha_{xy}$ increases with  $T$,
which saturates towards ln2 at the thermodynamic and large $T$
limit according to their saturated entropy per LL orbital.
In the insets of (a-b) we show the same date in the logarithmic plot, where
 a powerlaw scaling  is found $\alpha_{xy}\propto T^{\eta}$ for a range of  $T$.
(a) For $n=0$ LL with Coulomb interaction, $\eta\sim 0.54\pm 0.03$.
(b) For $n=1$ LL with Coulomb interaction, $\eta\sim 0.44\pm 0.03$.
The error bar is estimated by fitting low $T$ data from different $N_s$.
\label{fig:nu12}}
\end{figure}

The many-body Hamiltonian can be written as
\begin{eqnarray}
H&=&\sum _{i<j} \sum _{ \bf {q}}  e^{-q^2/2 }  V(q)
e^{i {\bf q} \cdot ({\bf R}_i -{\bf R}_j)} \nonumber \\
&+&\sum _{i} \sum _{ \bf q} e^{-q^2/4} U_{\bf q}
e^{i{\bf q }\cdot {\bf R}_i},
\end{eqnarray}
where ${\bf R}_i$ is the guiding center coordinate of the $i$-th
electron, $V(q)=2\pi e^2/\epsilon q$ is the
Coulomb potential, and $U_{\mathbf q}$ is the impurity potential
with the wave vector ${\bf q}$.
We set the magnetic length $\ell=1$ and  $e^2/\epsilon\ell=1$ for convenience.
The Gaussian white noise potential we use is generated according to the
correlation relation
in $q$-space
$\langle U_{\bf q}U_{{\bf q}'}\rangle=(W^2/A) \delta_{{\bf q},-{\bf q'}}$,
which corresponds to $\langle U({\bf r})U({\bf r'})\rangle
=W^2 \delta (\bf {r-r'})$\cite{Sheng2003}
in real space, where $W$ is the strength of the disorder and $A$ is the area of
the system.
The filling fraction is then defined as
$\nu=N_e/N_s$, where $N_e$ and $N_s$ are the number of electrons in the partially filled LL and
the number of the flux quanta, respectively.

Thermoelectric  conductivity $\alpha_{ij}$ is defined by the  electrical current generated by a temperature gradient in the absence of any voltage (short-circuit condition), or via Onsager relation, by the heat current generated by a voltage difference at a uniform  temperature. Since heat current is carried by thermal excitations, thermoelectric conductivity is purely a property of the partially occupied LL. In our case, its value depends on temperature $k_B T$ and disorder strength $W$ (both in unit of $e^2/\epsilon \ell$). While thermoelectric conductivity is conceptually convenient for theoretical analysis,  experiments usually measure thermopower $S_{xx}$ and Nernst signal $S_{xy}$ directly. These are given by the product of $\alpha_{ij}$ and resistivity $\rho_{jk}$: $S_{ik}= \alpha_{ij} \rho_{jk}$.

We first consider the clean limit $W=0$. As shown explicitly for both non-interacting systems \cite{Girvin1982, Fu} and generic interacting electron fluid\cite{Cooper1997, Yang2009}, in the absence of disorder, thermoelectric Hall conductivity $\alpha_{xy}$ is directly proportional to entropy density $s$: $\alpha_{xy} = s / B$. This remarkable formula enables us to obtain $\alpha_{xy}$ by numerically calculating entropy---a thermodynamic property---without invoking the Kubo formula for transport coefficients.

We perform thermal entropy calculations based on exact calculation  of energy spectrum of the  Hamiltonian, and obtain
$\alpha_{xy}=S/N_s$ (in units of
$k_Be/h$), where {\bf $S=sA$ } is the thermal entropy of the system.
We consider systems  with  sub-dimension of Hilbert space upto   $N_h= 102348$ ($2119036$)  for full (partial) diagonalizations,
 which is  slightly smaller than  the sizes used in ground state simulations and gives reliable results for all temperature regime
we considered.

\begin{figure}[t]
\includegraphics[width=1.7\linewidth]{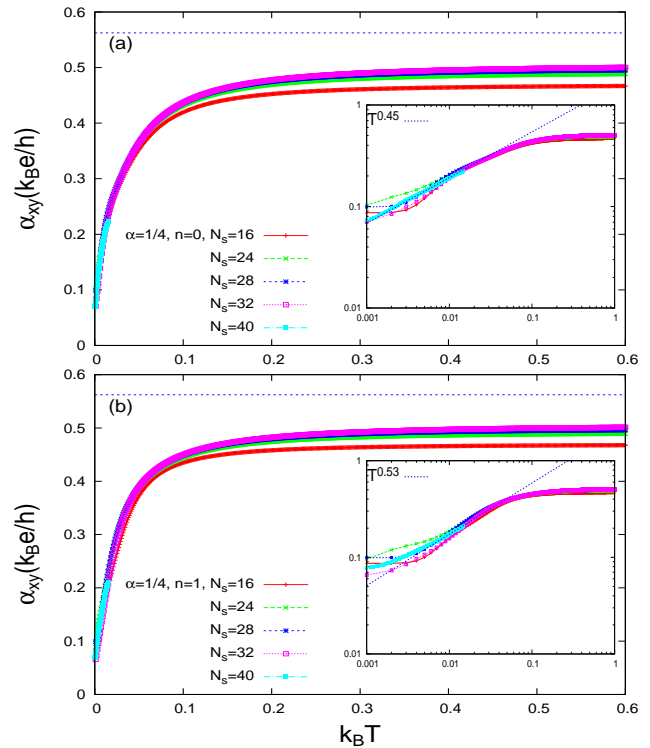}\\
\caption{$\alpha_{xy}$   for electrons
 at $\nu=1/4$ filling of the n-th LL with
 $N_s=16-40$.
 The dotted line above the data curve indicates the
expected $\alpha_{xy}$ at the infinite $T$ and $N_s$ limit according to their saturated entropy.
 The fitting shown in the insets of (a) and (b) illustrate  the powerlaw scaling  $\alpha_{xy}\propto T^{\eta}$.
(a) For $n=0$ LL with Coulomb interaction, $\eta\sim 0.45\pm 0.03$.
(b) For $n=1$ LL with Coulomb interaction, $\eta\sim 0.53\pm 0.03$.
\label{fig:nu14}}
\end{figure}


{\it Thermoelectric Hall response of composite Fermion liquids---} We first consider interacting quantum Hall systems without disorder scattering.
Two even denominator filling numbers $\nu=1/2$ and $1/4$ will be considered first, where low temperature
behavior of such systems is controlled by the physics of the composite Fermi-liquid\cite{Halperin1993}.
By exact diagonalization,  we can study  systems with upto the  number of flux quanta  $N_s=28$ 
for electrons at $1/2$ filling  using magnetic translational symmetry.
By obtaining all energy eigenvalues of the system, we determine $\alpha_{xy}$ from the entropy per flux.
As shown in Fig. 1(a-b), we show  $\alpha_{xy}$ for different  system sizes with
$N_s=12$ to $28$ for both $n=0$ and $1$ LLs.
The  $\alpha_{xy}$ grows with $T$ monotonically, and saturates
  towards  a universal value
$ln(2)*k_Be/h$  determined by the entropy per flux for LL at half-filling at high $T$ regime
with a small correction  due to the constraint of using canonical ensemble
with fixed particle number for finite size systems.   
As shown in insets of Fig. 1(a-b),
we identify a powerlaw behavior at low temperature $\alpha_{xy}\propto T^{\eta}$ as a straight line fitting to the data in the algorithmic plots, 
and find  the exponent  $\eta\sim 0.54\pm 0.03$   and $\sim 0.44\pm 0.03$ for $n=0$ and $1$, respectively.
We remark that although the nature of groundstates for the $n=0$ and $n=1$ are different at $T=0$  limit corresponding to
the composite-Fermi liquid and Moore-Read non-Abelian FQH state\cite{Moore1991}, respectively,
 $\alpha_{xy}$ of both systems in the small  to intermediate $T$  can demonstrate
similar scaling behavior. This is because $\alpha_{xy}$ is controlled by composite-Fermi liquid behavior
once $k_BT$ is comparable to the excitation gap of the FQH of $n=1$ LL.
 The exponent $\eta$ appears to be slightly different for these two cases.
Furthermore,  we compare  $\alpha_{xy}$ of  systems with different type of electron-electron interactions including
  the Haldane short-range pseudopotential, and  find quantitative similar results.
Importantly, we have demonstrated that  $\alpha_{xy}$ at low $T$ behaves distinctly different from the linear $T$ scaling of the Fermi liquid behavior.
The finite size effect only occurs for very low $T\lesssim 0.01$, which is expected due to finite energy level spacing for such systems.

Now we move on to  $\nu=1/4$ filling number, where  larger systems can be accessed with $N_s$ up to $32$ ($40$) for full (partial) diagonalization.
As shown in Fig. 2(a-b) for systems with  $N_s=16$ to $40$ for both  LLs $n=0$ and $n=1$,
$\alpha_{xy}$  increases with $T$ rapidly and it  saturates  to a universal value  ($0.562$) determined by the entropy per flux for the $1/4$ partially filled  LL  at high $T$ limit.  As shown in the insets of  Fig. 2(a-b),  we demonstrate
 a powerlaw behavior at low temperature $\alpha_{xy}\propto T^{\eta}$, where almost all data points from low $T$
 regime  can be well fitted by such a scaling behavior up to $T\sim 0.07$, beyond which  $\alpha_{xy}$  starts to saturate.
Clearly,  the finite size effect is reduced comparing to  $\nu=1/2$ case as we can access larger systems with larger $N_s$ at $\nu=1/4$ filling.
The exponent is identified to be  $\eta\sim 0.45\pm 0.03$   and $\sim 0.53\pm 0.03$ for $n=0$ and $1$, respectively.
The average exponent for the powerlaw behavior is consistent with $\eta \sim 0.50$ for both  $\nu=1/4$ and $1/2$ filling numbers, indicating a possible universal scaling behavior for $\alpha_{xy}$ at low $T$.

\begin{figure}[t]
\includegraphics[width=0.74\textwidth]{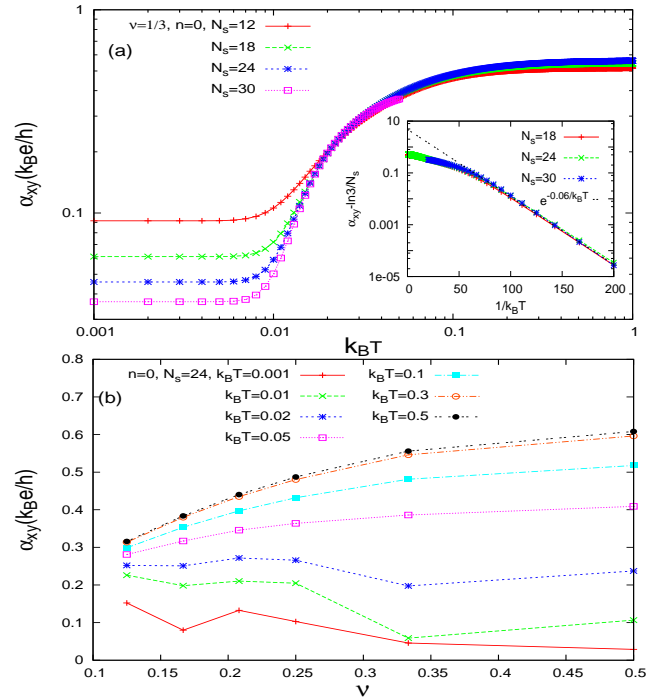}\\
\caption{$\alpha_{xy}$   for electrons
 at different filling numbers of the lowest  LL for
pure  interacting systems. (a)
$\alpha_{xy}$ vs. $k_BT$ for  $N_s=12-30$ at fixed $1/3$ filling number.  In the inset we show
$\alpha_{xy}-ln3/N_s$ as a function of $1/k_BT$ for larger systems $N_s=18-30$. (b)
$\alpha_{xy}$ vs. $\nu$ for  $N_s=24$ for different   $k_BT=0.001$ to $0.5$.
}\label{fig:nu13}
\end{figure}

\begin{figure}[t]
\includegraphics[width=1.3\linewidth]{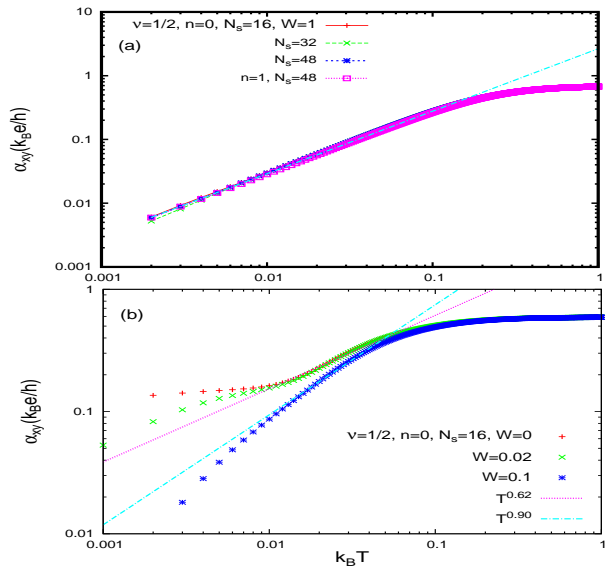} \\
\caption{(a) $\alpha_{xy}$
 for {\it noninteracting} electrons at half-filling of the $n=0$ and $1$  LLs  with
 $N_s=16-48$ and    the disorder strength $W=1$. The dashline shows the linear fit of the data.
(b) $\alpha_{xy}$   for electrons
 at half-filling of the lowest  LL for
disordered interacting systems  with  $N_s=16$.
\label{fig:disorder}}
\end{figure}


{\it Thermoelectric Hall response for different electron filling numbers---}
At low temperature, correlated states emerge for interacting quantum Hall systems.  The nature of the states
dependent on the electron filling number.  As we have shown before, the even denominator state is either
gapless (for $n=0$ case) or having a small excitation gap for QHE state ($1/2$ filling of $n=1$ LL),  which exhibits powerlaw
scaling for $\alpha_{xy}$ down to very low temperature.
 To explore possible distinct physics from thermoelectric Hall effect of a
FQH state with a robust gap,   we present results  of 1/3 FQH with Coulomb interaction for pure system at $W=0$.
As shown in Fig. 3(a) with flux quanta $N_s=12-30$, we find that  $\alpha_{xy}$ at low $T<0.01$ decreases  with the increase  of
$N_s$, while $N_s\alpha_{xy}=ln3$ in accordance with the three fold degeneracy of the system.
The data for the largest system $N_s=30$ is obtained through Lanczos for lowest 600 states in each momentum sector,
which allows us to  obtain lower temperature $\alpha_{xy}$ accurately.
The temperature regime to see such a vanishing $\alpha_{xy}$ behavior is $T\leq 0.01$, indicating a finite
gap for such a   fractionalized topological state.
Beyond $T>0.01$,   ${\alpha}_{xy}$ increases with $T$
very sharply, and saturates to a universal value at larger $T$ side.
To compare these data  with the activation behavior of a gapped  system,  we
use the following form $\alpha_{xy}-ln3/N_s \propto exp(-\frac {E_g} {k_BT})$ to fit the low $T$ data.
The constant term $ln3/N_s$ is the size dependent contribution  from topological degeneracy.
As  shown in the inset of Fig. 3(a),  we  identify the collective excitation gap $E_g=0.06$, which should be distinguished from the quasiparticle charge gap. Therefore thermoelectric response provides a new
experimental way to detect collective excitations.

We compare  ${\alpha_{xy}}$ at other filling numbers with the behavior of composite Fermion liquids systems at $\nu=1/2$ and $1/4$.
In Fig. 3(b), we show $\alpha_{xy}$ vs. $\nu$ for filling numbers $\nu=1/8-1/2$ at fixed $N_s=24$
(the result is symmetric about $\nu=1/2$ due to particle-hole symmetry).
At low temperature,  the gapped 1/3  FQH state has suppressed  $\alpha_{xy}$ as discussed above.
Other systems are either gapless,  or having  tiny gaps comparing to 1/3 FQH system.
We find clearly local minimum at $\nu=1/3$ for $k_BT \leq 0.02$.
As $T$ is further increased,  we identify $\alpha_{xy}$  as a smooth increasing function of $\nu$,
  which reaches a  maximum value at    $\nu=1/2$.
 $\alpha_{xy}$ also saturates with $T$ to different values in accordance with the maximum entropy  at different $\nu$.
  We have  further examined the scaling behavior of $\alpha_{xy}$ at low
filling numbers $\nu=1/6$ and $1/8$,  where a Wigner solid may emerge at low temperature limit. At finite $T$,  we
identify a powerlaw scaling of  $\alpha_{xy}\propto T^{\eta}$,  with $\eta\sim 0.4\pm0.04$ for larger system sizes $N_s=32-36$.
We caution that  $\eta$ identified here shows a trend of increasing with $N_s$ at the largest system we can access.
However,  the powerlaw behavior appears to be robust for low $T$ regime for such systems at low filling numbers.

{\it Disorder effect---}
For experimentally realized  quantum Hall systems,  impurity scattering effect always presents
in additional to electron-electron  Coulomb interaction,   thus we need to include
both terms  in the Hamiltonian  to model realistic systems.
We first show  the  results of $\alpha_{xy}$ (entropy) for  disordered quantum Hall systems  without considering
Coulomb interaction.
In this case, the Hilbert space has a dimension
 $N_s$,  and the obtained $\alpha_{xy}$ as a function of  $k_BT$
is shown in
Fig. 4(a).  For system flux number $N_s=16-48$, we find that all the data collapse into one universal curve as a function of $T$.
The low $T$ results can be well fitted by  
a linear dependence  $\alpha_{xy}\propto T$
 as shown as the dashed line fitting in the log-log plot in Fig. 4(a), which is  independent of the system sizes $N_s$.
The linear $T$ dependence of $\alpha_{xy}$ is consistent with results of other noninteracting systems
with strong magnetic field\cite{Girvin1982} as also  obtained for graphene systems based on Kubo formula\cite{Zhu2010}.
The $\alpha_{xy}$ saturates
to an universal value $ln2*k_Be/h$ as $T$ approaches the order of 1.

Now we turn to the effect of weak random disorder for interacting quantum Hall systems.
In this case, the magnetic translational symmetry is broken
due to momentum nonconserving scattering present in the Hamiltonian.  So we have to diagonalize the whole Hilbert space
for $N_e$ electrons occupying $N_s$ orbitals.   This limits us to smaller systems with $N_s=16$  and $18$.
 As shown in Fig. 4(b) for $N_s=16$ at filling number $\nu=1/2$,  thermal entropy is always an increasing function of $T$ for different $W$,
which  saturates to the  universal value determined by the maximum entropy per orbital in such systems.
Entropy decreases monotonically at small $T$ regime with the increase of $W$.
In the smaller and intermediate $T$ regime,  entropy appears to follow the
 powerlaw scaling behavior $s \propto T^{\eta}$,  with the exponent $\eta$ increases from around $0.62$
for $W=0.02$, to $\eta\sim 0.90$ for $W=0.1$. Very similar results are obtained for $N_s=16$ and $18$,
indicating these behaviors are robust as they are associated with the thermal entropy.
Quantitatively,  the exponents determined here are less reliable due to limited system sizes we can access.

{\it Summary and discussion.---} We now discuss the importance of our work for understanding the thermoelectric Hall effect
of different quantum Hall systems at low temperature. Since most of quantum Hall systems realized in experiments have high mobility,  the interaction
effect plays essential role in lifting LL degeneracy.  Our work focuses on finite temperature,
where these interacting systems are either in gapless composite Fermi-liquid states,  or  thermally  excited QHE states.
The nonlinear power law scaling  behavior we established for $\nu=1/2$ and $1/4$ filling numbers strongly proves that these systems
at finite  temperature  are strongly correlated electron fluids, distinctly different from Fermi liquids for which the linear scaling law of $\alpha_{xy}$ is expected from the low energy excitations around the Fermi level.
The scaling behavior of $\alpha_{xy}\propto  T^{\eta}$ (with $\eta\sim 0.5$) appears to be very general for quantum Hall systems at different filling numbers.
In particular,  the $\nu=1/2$ and $1/4$ quantum Hall states have  much enhanced $\alpha_{xy}$ at the lowest temperature, and therefore are promising candidates for
thermoelectric energy conversion with high efficiency\cite{Fu}.
Our calculations can be naturally extended to other quantum Hall systems\cite{Heremans2017,Zeng2019,Veyrat2019,Gusev2017,Sato2019,Barlas2012}    with different LL degeneracy or multi-component interactions such as graphene.  The intriguing strange metal transport we discovered  calls for a  theory of fractional quantum Hall liquids at finite temperature, which we will provide in forthcoming works.

\textit{Acknowledgments.---}
This work was supported by the U.S. Department of Energy  (DOE),  Office of Basic Energy Sciences under the grant No. DE-FG02-06ER46305 (D.N.S) and  by DOE Office of Basic Energy Sciences under Award DE-SC0018945 (L.F.). LF was supported in part by the David and Lucile Packard Foundation.

\end{document}